\documentclass{article}
\usepackage{epsf}
\usepackage{dcolumn}
\usepackage{array}
\newcommand{\ap}[3]{     {\it Ann. Phys. (NY) }{\bf{#1},} (19#3) #2}
\newcommand{\jmp}[3]{   {\it J. Math. Phys. } {\bf{#1},}  (#3) #2 }

\newcommand{\mpla}[3]{    {\it Mod. Phys. Lett. }{ A \bf{#1},} (19#3) #2 }
\newcommand{\pla}[3]{    {\it Phys. Lett. }{ A \bf{#1},} (#3) #2 }
\newcommand{\prd}[3]{    {\it Phys. Rev. }{ D \bf{#1},} (19#3) #2 }

\newcommand{\prl}[3]{    {\it Phys. Rev. Lett. }{\bf{#1},} (19#3) #2 }

\newcommand{\eq}[1]{{Eq.~(\ref{#1})}}

\newcommand{\etal}{{\it et al.}}
\newcommand{\ie}{{\it i.e. }}

\newcommand{\bea}{\begin{eqnarray}}
\newcommand{\beq}{\begin{equation}}
\newcommand{\eea}{\end{eqnarray}}
\newcommand{\eeq}{\end{equation}}
\newcommand{\nnu}{\nonumber}
\begin{document}
\title{Symmetric Triple Well with Non-Equivalent Vacua: Instantonic Approach }
{\small
\author{H. A. Alhendi$^2$\ and \ E. I. Lashin$^{1,2,3}$\\
$^1$ The Abdus Salam ICTP, P.O. Box 586, 34100 Trieste, Italy\\
$^2$ Department of Physics and Astronomy, College of Science,\\
King Saud University, Riyadh,
Saudi Arabia \\
$^3$ Department of Physics, Faculty of Science, \\Ain Shams
University, Cairo, Egypt} } \maketitle
\begin{abstract}
We show that for the triple well potential with non-equivalent
vacua, instantons generate for the low lying energy states a
singlet and a doublet of states rather than a triplet of equal
energy spacing. Our energy splitting formulae are also confirmed
numerically. This splitting  property is due to the presence of
non-equivalent
vacua. A comment on its generality to multi-well is presented.
\end{abstract}

Instantons are non-trivial classical solutions of Euclidean field
equations for which the action is finite \cite{hoft}. Their
importance, besides being topological configurations, comes from
their finite contributions to Feynmann path integral. In quantum
mechanics instantons correspond to non-trivial finite classical
solutions of classical equations of motion with inverted potential
\cite{col}.

The uses of instanton calculations have proven to be useful in
analyzing non-perturbative aspects of quantum mechanical systems
with degenerate vacua, this is because instanton solutions
contribute to the quantum tunnelling phenomena, which can be
calculated with the aid of the dilute gas approximation. A
celebrated example is the splitting of energy level in the
symmetric double well case \cite{col}.

To our knowledge only a few recent papers attempted to apply  the
instanton method to the triple well potential with non-equivalent
vacua \cite{lee,casa,sato}. This problem is rather involved
compared to the symmetric double well case.  It has been suspected
that in the presence of non-equivalent vacua the dilute gas
approximation may break down \cite{sato}. Moreover it has been
claimed\cite{lee,casa} that the average of the harmonic
frequencies over the non-equivalent vacua of the potential serves
as the central position for the equidistant nearly degenerate
first three levels. Unfortunately this claim does not lead to the
correct spectra and is in contradiction with the numerical
calculations. Thus understanding the structure of the energy
levels in a triple well becomes important. These issues encourage
us to look at the problem more carefully and to our knowledge a
proper treatment has not been previously carried out.

In this work, we consider the triple well potential of the form,
which admits inversion symmetry $(x\rightarrow -x)$, \beq
V(x)={\omega^2 \over 2}\, x^2\,(x^2-1)^2, \label{pot} \eeq
\begin{figure}[htbp]
\epsfxsize=7cm
\centerline{\epsfbox{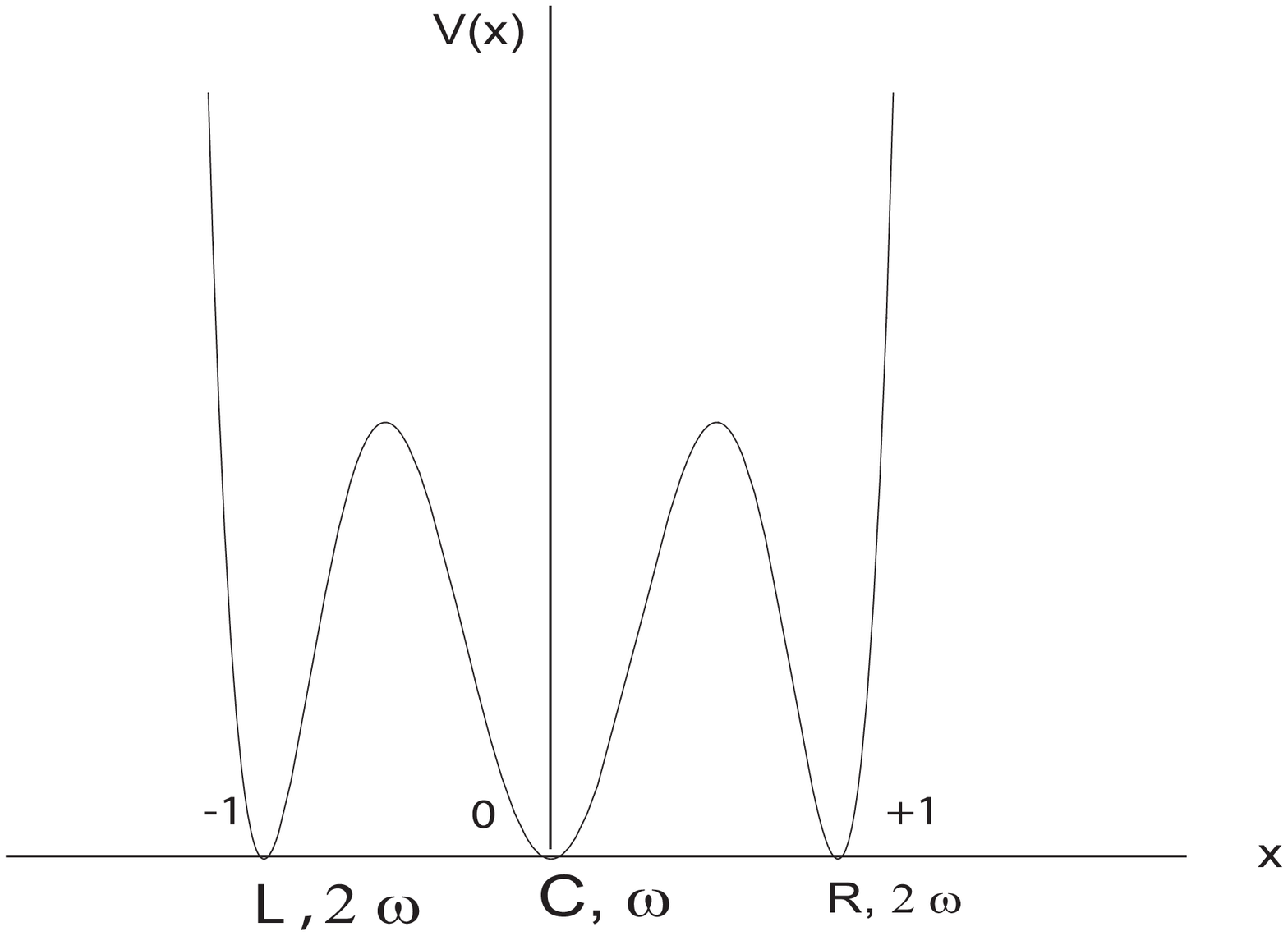}}
\caption{\footnotesize
Triple well: $V(x)={\omega^2 \over 2}\, x^2\left(x^2-1\right)^2$}
\label{triplewell}
\end{figure}
The potential ,$V(x)$, shown in Fig.~\ref{triplewell} consists of
central, left and right well separated by barriers. In the
vicinity of each well (near the minima at $x=0,\pm 1$)
the potential can be approximated as a harmonic oscillator potential.
The frequency corresponding to the
central well is $\omega_0=\omega$, while for the left and right well
is $\omega_1=2\,\omega$
as can be easily verified by expanding the potential around each
minima. These different frequencies show the different curvatures at each minima.

Computation of transition probability amplitudes between different
minima allows to extract the low lying energy eigenvalues.
Here the instanton contributions to these transitions are calculated by making a proper
treatment for the non-equivalent vacua. As will be shown this proper treatment
has a drastic effect on the energy spectra in contrast with what has been claimed.

For the potential in \eq{pot} we have four kinds of instanton
solutions connecting neighboring  minima and have the forms:
\bea
x^{I}(t)=\pm {1\over
\left[1+e^{\mp\omega\left(t-t_0\right)}\right]^{{1\over 2}}}
&,&
x^{\bar{I}}(t)=\pm {1\over
\left[1+e^{\pm\omega\left(t-t_0\right)}\right]^{{1\over 2}}},
\nnu \\
\label{inssoln}
\eea
where $I(\bar{I})$ indicates instanton(anti-instanton)
respectively.

The transition from the  minimum at $x=0$ to that at $x=1$ during
the Euclidean time interval $T$ can be written as:
\beq
\langle
1|e^{-H T}|0\rangle = \langle 1|E_0\rangle\,\langle
E_0|0\rangle\;e^{-E_0\,T} + \langle 1|E_2\rangle\,\langle
E_2|0\rangle\;e^{-E_2\,T}\;+\cdots
\label{tr01}
\eeq
Due to the symmetry only even states contribute (odd states are vanishing at
$x=0$). On the other hand for the transition from $x=1$ to $x=1$,
we have
\beq
\langle 1|e^{-H T}|1\rangle = |\langle
1|E_0\rangle|^2\;e^{-E_0\,T} + |\langle
1|E_1\rangle|^2\;e^{-E_1\,T} + |\langle
1|E_2\rangle|^2\;e^{-E_2\,T}\;+\cdots
\label{tr11}
\eeq
These transitions, Eqs.~(\ref{tr01})-(\ref{tr11}), are sufficient to
extract the low lying energy eigenvalues namely $E_0, E_1$ and
$E_2$.

\begin{figure}[htbp]
\epsfxsize=7cm
\centerline{\epsfbox{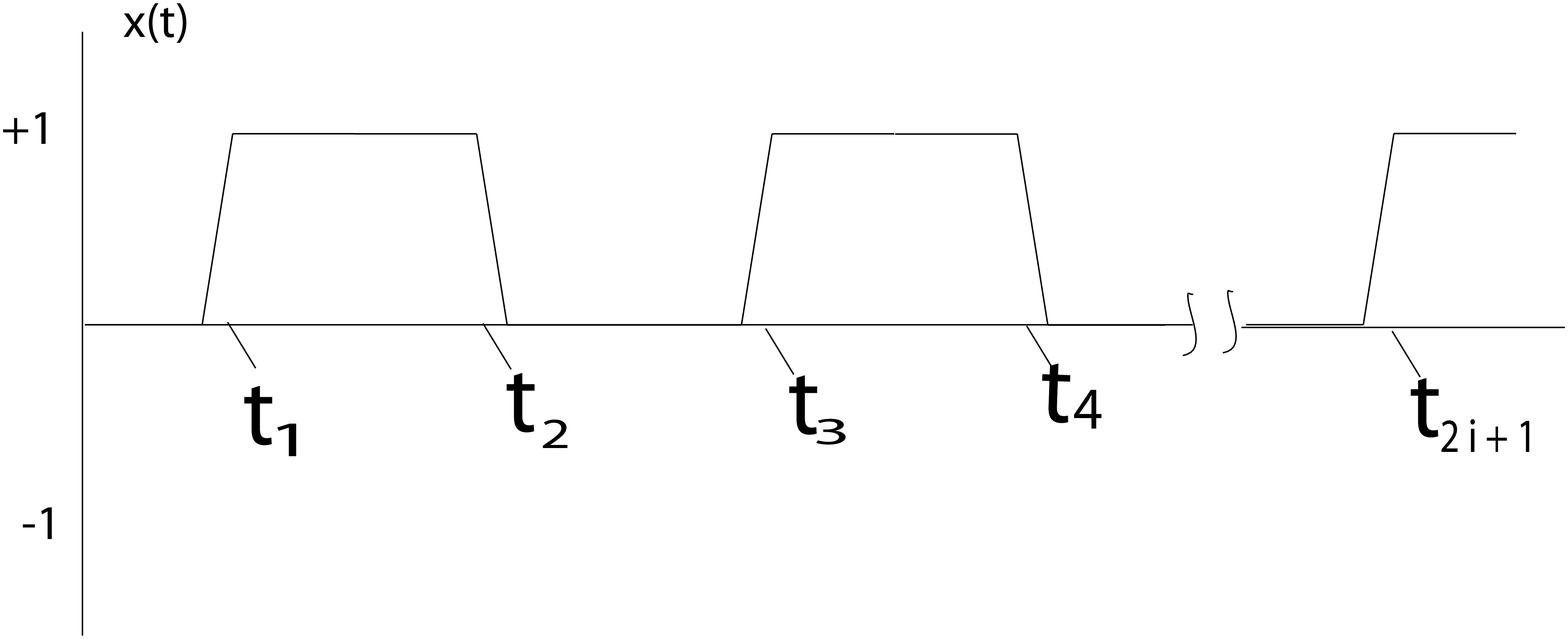}}
\caption{\footnotesize
The multi instanton contribution to the transition from $x=0$ to $x=1$.
The solutions with positive slope are called instantons whereas those
of negative are called anti-instanton}
\label{odd}
\end{figure}

In the dilute gas approximation, a typical instantonic
contribution to the transition in \eq{tr01} is found to be
composed of $i+1$ instantons and $i$ anti-instanton as shown in
Fig.~\ref{odd} which  we call, for later use, $M_i^{\mbox{odd}}$.
It should be stressed that this contribution has $2^i$ different configurations which should
be taken into consideration during calculations.

\begin{figure}[htbp]
\epsfxsize=7cm
\centerline{\epsfbox{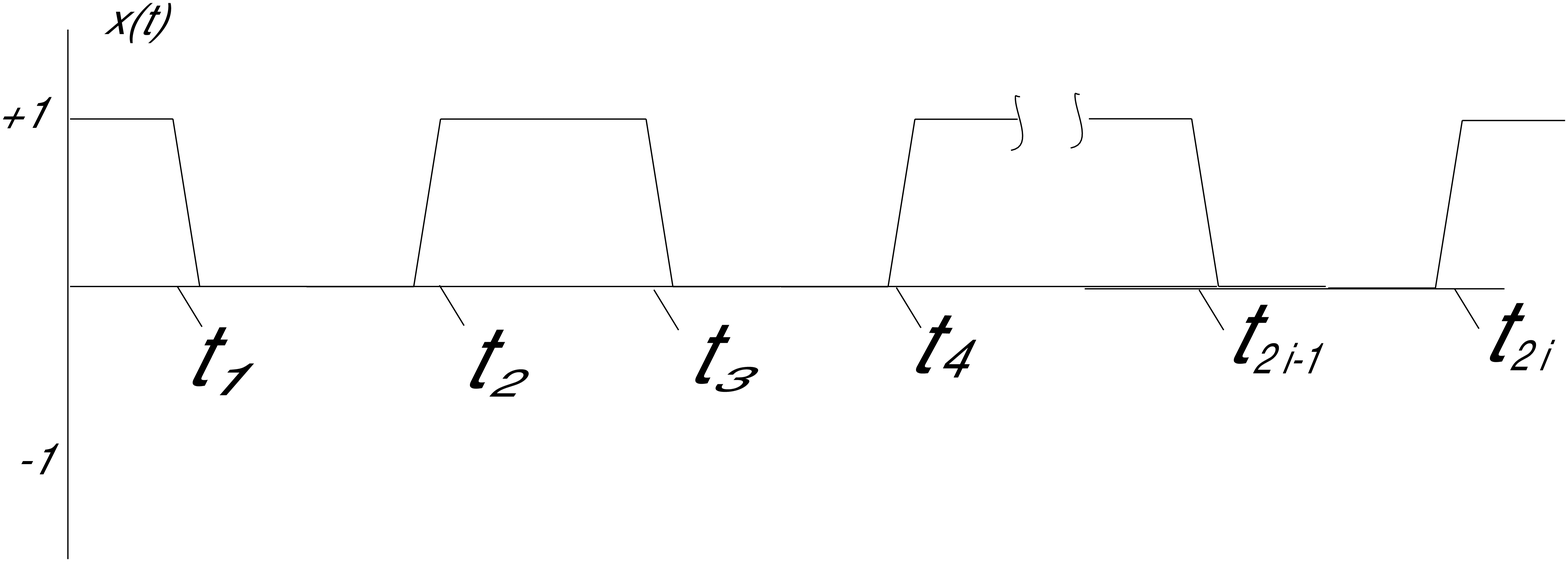}}
\caption{\footnotesize
The multi instanton contribution to the transition from $x=1$ to $x=1$.
The solutions with positive slope are called instantons whereas those
of negative are called anti-instanton}
\label{even}
\end{figure}

For the case of the transition, \eq{tr11},  one finds that it is composed of $i-$instantons and
$i$ anti-instantons as shown in Fig.~\ref{even} which we call $M_i^{\mbox{even}}$. This
contribution has $2^{i-1}$ different configurations. Also there is a contribution coming
from the trivial solution $(x(t)=1)$ that should be taken into consideration.

It is worthy to note the following remarks. Firstly, in evaluating the transition drawn in
Figs.~(\ref{odd},\ref{even}) we have to compute the fluctuations over
these paths consisting of strings of instantons-anti-instantons.
However, because instantons are quite localized (with extension of order
$\omega^{-1}$), we can neglect the fluctuations around their
bodies. In other words, the quantum fluctuations would be
effective only around the straight line paths in
Figs.~(\ref{odd},\ref{even}). Along each step of the straight line
paths, the particle resides at the minima of the potential
(maxima for the inverted one) shown in Fig.~\ref{pot}. These
fluctuations can be treated like the case of the harmonic
oscillator. The  crucial  point here is to take care that the frequencies corresponding
to each step may be different and can not be taken to be equal.
Secondly, to compute the transition probability amplitudes given by Eqs.~(\ref{tr01})-(\ref{tr11}) one
should sum over all the possible multi-instanton-anti-instanton configurations. An elegant method for
avoiding this complicated sum has been developed and implemented in Refs.~\cite{rossi1,rossi2} to
the case of quantum mechanical system with degenerate classical minima (exemplified by the symmetric double
well) giving correct results for the low lying energy spectra and also non trivial information about the
wave functions. The method in Refs.~\cite{rossi1,rossi2} is based on saturating the transition
probability amplitudes from the minimum to itself by the trivial solution, while for the transitions between
two successive minima by one-instanton.
However to extract energy eigenvalues one should have a priori some knowledge about the pattern of the correct
energy spectra. Since in our present work we do not have such a priori knowledge about the correct pattern for
the energy spectra for the symmetric triple well with non-equivalent vacua,
thus the sum over the multi-instanton-anti-instanton configurations can not be avoided.

Using a technique similar to that used for the double well with
non-equivalent vacua \cite{rivero}, we rearrange the n-instanton
integrals in such a form that recursive relations can be given for
all these integrals. With the help of these relations the
instanton sum can be easily managed. As stated before, only two
transitions are required to be calculated, namely that between the
minima at $x=0$ and $x=1$ and the other between $x=1$ to $x=1$.

The contribution of the $M_i^{\mbox{odd}}$  has the integral:
\bea
M_i^{\mbox{odd}} &=& 2^i\,N K^{2\,i+1} A^{2\,i+1} \int_{-{T\over
2}}^{{T\over 2}} dt_1\;\int_{t_1}^{{T\over 2}} dt_2\cdots\cdots
\int_{t_{2\,i}}^{{T\over 2}}\,dt_{2\,i+1}\nnu\\
&&e^{-{1\over 2}\omega_0(t_1-(-{T\over 2}))}\;e^{-{1\over 2}\omega_1(t_2-t_1)}\;
e^{-{1\over 2}\omega_0(t_3-t_2)}\cdots
e^{-{1\over2}\omega_1({T\over 2}-t_{2\,i+1})},\nnu\\
&=& 2^i N K^{2\,i+1} A^{2\,i+1}\;e^{-{1\over 2}{\omega_0+\omega_1\over 2}T}\; \int_{-{T\over
2}}^{{T\over 2}} dt_1\;\int_{t_1}^{{T\over 2}} dt_2\cdots\cdots
\int_{t_{2\,i}}^{{T\over 2}}\,dt_{2\,i+1}\nnu\\
&&e^{\delta\,t_1}\,e^{-\delta\,t_2}\,e^{\delta\,t_3}\cdots
e^{\delta\,t_{2\,i+1}},
\nnu\\
\label{tr1}
\eea
where $(\,N = \sqrt{{\omega \over 2\,\pi}}\,)$ is
the normalization constant, $A$ is the exponential of the
classical Euclidean action for one instanton and $\delta={\omega_1
- \omega_0\over 2}$. The factor $K$ is calculated by matching the
one instanton contribution. The one instanton contribution for the
potential ,\eq{pot}, can be found in Refs.~\cite{lee,casa}, but
with paying attention  to used different conventions.

Now this integral can be put in an equivalent form:
\beq
M_i^{\mbox{odd}}={N\over \sqrt{2}}\,
B^{2\,i+1}\,e^{-{1\over 2}{\omega_0+\omega_1\over 2}T}\;\int_{-{T\over 2}}^{{T\over 2}} \;dt
\;e^{\delta t} \,{({T\over 2} +t)^i \over i!}\, {({T\over 2}-t)^i \over i!},
\label{oddins}
\eeq
where $B=\sqrt{2}\, K\,A$ .

For the transition from $x=1$ to $x=1$, the corresponding integral
is
\bea
M_i^{\mbox{even}} &=& 2^{i-1}\,N K^{2\,i} A^{2\,i}
\int_{-{T\over 2}}^{{T\over 2}} dt_1\;\int_{t_1}^{{T\over 2}}
dt_2\cdots
\cdots \int_{t_{2\,i-1}}^{{T\over 2}}\,dt_{2\,i} \nnu \\
&&e^{-{1\over 2}\omega_1(t_1-(-{T\over 2}))}\;e^{-{1\over 2}\omega_0(t_2-t_1)}\;
e^{-{1\over 2}\omega_1(t_3-t_2)}\cdots
e^{-{1\over2}\omega_1({T\over 2}-t_{2\,i})},\nnu\\
&=& N 2^{i-1}\,K^{2\,i} A^{2\,i}\; e^{-{\omega_1 \over 2} T}\; \int_{-{T\over 2}}^{{T\over 2}}
dt_1\;\int_{t_1}^{{T\over 2}} dt_2\cdots
\cdots \int_{t_{2\,i-1}}^{{T\over 2}}\,dt_{2\,i} \nnu \\
&&e^{-\delta\,t_1}\,e^{\delta\,t_2}\,e^{-\delta\,t_3}\cdots
e^{\delta\,t_{2\,i}},
\nnu\\
\label{tr2}
\eea
which can also be put in an another equivalent form
\beq
M_i^{\mbox{even}}  = {N\over 2} B^{2\,i}\,e^{-{1\over
2}{\omega_0+\omega_1\over 2}T}\; \int_{-{T\over 2}}^{{T\over 2}}
\;dt\; e^{\delta t} \;{({T\over 2} +t)^{i-1} \over (i-1)!}\,
{({T\over 2}- t)^i \over i!}.
\label{evenins}
\eeq

Carrying out the sum over the multi-instantons is rather involved,
but can be simplified by studying the basic integral of the form
\beq
I(n,m)= B^{n+m+1}\,
e^{-{1\over 2}{\omega_0+\omega_1\over 2}T}\;\int_{-{T\over 2}}^{{T\over 2}}
\;dt \;e^{\delta t} \,{({T\over 2} +t)^n \over n!}\,{({T\over 2}- t)^m \over m!}.
\label{basicdf}
\eeq
Integrating by parts we get the following recursive relations
(here we drop the common factor $e^{-{1\over 2}{\omega_0+\omega_1\over 2}T}$
in the definition of $I(n,m)$)
\bea
I(n,0)&=&{B\over \delta}\,\left[{(B\,T)^n \over n!}\,
e^{\delta\,{T\over 2}}- I(n-1,0)\right],\nnu \\
I(0,m)&=&{B\over \delta}\,\left[I(0,m-1)- {(B\,T)^m \over m!}\,
e^{-\delta\,{T\over 2}}\right],\nnu \\
I(n,m)&=&{B\over \delta}\,\left[I(n,m-1)- I(n-1,m)\right].\nnu\\
\label{rec1}
\eea
With the help of these recursive relations,
Eqs.~(\ref{rec1}), and taking into account the correct counting
together with the starting initial integral $ I(0,0)={B\over
\delta}\,\left[ e^{\delta\,{T\over 2}} - e^{-\delta\,{T\over
2}}\right], $ we get (here we keep the common factor in the
definition of $I(n,m)$) \bea I(n,m)&=&e^{-{1\over 2}
\omega_0\,T}\sum_{i=0}^{n} \left(
\begin{array}{c}
m+n-i\\
m
\end{array}
\right)
(-1)^{n-i}\;\left({B\over\delta}\right)^{n+m-i+1}\,{(B\,T)^i \over
i\,!} \nnu \\
 &+ & e^{-{1\over 2} \omega_1\,T}\sum_{j=0}^{m}
\left(
\begin{array}{c}
m+n-j\\
n
\end{array}
\right)
(-1)^{n+1}\;\left({B\over\delta}\right)^{n+m-j+1}\,{(B\,T)^j \over
j\,!}, \nnu \\
\label{bas2} \eea where $(^{\,m}_{\,n} )$ are the binomial
coefficients.

For the transition from the minimum at $x=0$ to the one at $x=1$
we need to evaluate the sum $\sum_{i=0}^{\infty}\,I(i,i)$ which is
still hard to perform using \eq{bas2}. To facilitate evaluating
the sum, we define $S(n,m)=\sum_{i=0}^{\infty}\,I(n+i,m+i)$ and
let $S^{\pm}_j(n,m)$ be the coefficient of $S(n,m)$ associated
with the terms ${(B\,T)^j \over j!}\,e^{\pm\,{\delta\,T\over 2}}$.
Using the recursion relations, Eqs.~(\ref{rec1}), one gets \bea
S_i^{\pm}(n,m)&=&{B\over
\delta}\left[S_i^{\pm}(n,m-1)-S_i^{\pm}(n-1,m)\right],\nnu\\
S_i^{+}(n,m)&=& S_{i+1}^{+}(n+1,m),\nnu \\
S_i^{-}(n,m)&=& S_{i+1}^{-}(n,m+1),\nnu \\
S_i^{+}(n,m)&=& S_{i}^{+}(i,m+i-n)\;\;\; \mbox{where n $<$ i} ,\nnu \\
S_i^{-}(n,m)&=& S_{i}^{-}(n+i-m,i)\;\;\; \mbox{where m $<$ i} .\nnu \\
\label{rec2}
\eea
By letting $a_{i}^{\pm}\equiv S_{i}^{\pm}(i,i)$,
then with the aid of the recursion relations, Eqs.(\ref{rec2}), we
get \beq a_{i+1}^{\pm}=a_{i-1}^{\pm}\mp{\delta\over
B}\,a_{i}^{\pm}, \label{rec3} \eeq consequently the sum over the
multi-instantons becomes \beq
\sum_{i=0}^{\infty}\,M_i^{\mbox{odd}} ={N\over
\sqrt{2}}\,\sum_{i=0}^{\infty}\,I(i,i)= {N\over
\sqrt{2}}\,\sum_{i=0}^{\infty}\,\left(a^{+}_{i}\,{(B\,T)^i \over
i!}\,e^{-\,{\omega_0\,T\over 2}} + a^{-}_{i}\,{(B\,T)^i \over
i!}\,e^{-\,{\omega_1\,T\over 2}}\right). \label{multi1} \eeq The
coefficient $a^{+}_{i}$ corresponds to the coefficient of linear
combination of two exponential (note that the factorial factor is
removed) namely that \beq C_{+}\,e^{\alpha_+} +
C_{-}\,e^{\alpha_-}, \label{sol1} \eeq where $\alpha_{\pm}$ are
found to be \beq \alpha_{\pm}=-{\delta\over 2 B} \pm\,
\sqrt{\left({\delta\over 2 B}\right)^2+1}. \label{root1} \eeq

The coefficients $C_{+}$ and $C_{-}$ can be then determined from
\bea
a_0^{+}&=& C_{+}\,+C_{-}, \nnu \\
a_1^{+} &=& C_{+}\,\alpha_{+}\, + C_{-}\,\alpha_{-},\nnu\\
    &=& C_{+}\,\left(-{\delta\over 2 B} +\,
\sqrt{\left({\delta\over 2 B}\right)^2+1}\;\right)\nnu\\
&+&
C_{-}\,\left(-{\delta\over 2 B} -\,
\sqrt{\left({\delta\over 2 B}\right)^2+1}\;\right).\nnu\\
\label{sol2}
\eea

Thus from \eq{bas2} we get
\beq
a_{0}^{+}=\sum_{i=0}^{\infty}\,
\left(
\begin{array}{c}
2\,i\\
i
\end{array}
\right)
(-1)^{i}\,\left({B\over
\delta}\right)^{2\,i+1}={B\over\delta}{1\over \sqrt{1+\left({ 2
B\over \delta}\right)^2}} = {1\over 2 \sqrt{1+\left({\delta\over 2
B}\right)^2}},
\label{a0p}
\eeq
and
\beq
a_{1}^{+}=\sum_{i=1}^{\infty}\,
\left(
\begin{array}{c}
2\,i-1\\
i
\end{array}
\right) (-1)^{i-1}\,\left({B\over \delta}\right)^{2\,i} = {1\over 2} -{1\over 2}\,{1\over
\sqrt{1+\left({2 B\over \delta}\right)^2}}. \label{a1p} \eeq leading to  $C_{+}={1\over 2
\sqrt{1+\left({\delta\over 2 B}\right)^2}}$ and $C_{-}=0$. A similar treatment can be done for
$a_{i}^{-}$ leading to the results $C_{-}=-{1\over 2 \sqrt{1+\left({\delta\over 2 B}\right)^2}}$
and $C_{+}=0$. Using these results together with \eq{multi1}, one finally obtains for the odd
multi-instantons sum:
\beq
\sum_{i=0}^{\infty}\,M_i^{\mbox{odd}}= {N\over 2\,\sqrt{2}\,
\sqrt{1+\left({\delta\over 2 B}\right)^2}}\left[e^{-E_0\,T}-e^{-E_1\,T}\right], \label{fin1} \eeq
with \bea E_0&=& {(\omega_0 + \omega_1)\over 4}-\sqrt{{\delta^2\over 4} +
B^2},\nnu\\
E_2&=& {(\omega_0 + \omega_1)\over 4}+\sqrt{{\delta^2\over 4} +
B^2}. \nnu\\
\label{ener1}
\eea

Now the transition  probability amplitude  between the minimum at
$x=1$ to itself, after including the contribution of the trivial
solution $(x(t)=1)$, becomes
\beq
\langle 1|e^{-H\,T}|1\rangle=
N\,e^{-{\omega_1\,T\over 2}} +
\sum_{i=1}^{\infty}\,M_i^{\mbox{even}}=  N\,e^{-{\omega_1\,T\over
2}}+ {N\over 2}\, \sum_{i=1}^{\infty}\,I(i-1,i) ,
\label{multi3}
\eeq
which can be evaluated by the same procedure applied for the
transition  between the minimum at $x=0$ to that at $x=1$. For
this purpose we let $\Lambda_i^{\pm}=S_{i}^{\pm}(i,i+1)$ where
$S_{i}^{\pm}(i,i+1)$ defined as before. Using the recursion
relations in Eqs.~(\ref{rec2}) we obtain
\beq
\Lambda^{+}_{i}={B\over
\delta}\left[\Lambda^{+}_{i-1}-\Lambda^{+}_{i+1}\right],
\label{lamp}
\eeq
The  coefficient $\Lambda^{+}_{i}$ corresponds to the coefficient
of linear combination of two exponential (note that the factorial
factor is removed) namely that
\beq
D_{+}\,e^{\lambda_+} + D_{-}\,e^{\lambda_-} ,
\label{sold}
\eeq
where $\lambda_{\pm}$ are found to be
\beq
\lambda_{\pm}=-{\delta\over 2 B} \pm\, \sqrt{\left({\delta\over 2 B}\right)^2+1}.
\label{rootd1}
\eeq

The coefficients $D_{+}$ and $D_{-}$ can be then determined from
\bea
\Lambda_0^{+}&=& D_{+}\,+D_{-}, \nnu \\
\Lambda_1^{+} &=& D_{+}\,\lambda_{+}\, + D_{-}\,\lambda_{-},\nnu\\
    &=& D_{+}\,\left(-{\delta\over 2 B} +\,
\sqrt{\left({\delta\over 2 B}\right)^2+1}\;\right)\nnu\\
&+&
D_{-}\,\left(-{\delta\over 2 B} -\,
\sqrt{\left({\delta\over 2 B}\right)^2+1}\;\right).\nnu\\
\label{sol3}
\eea

Again using \eq{bas2} we get
\beq
\Lambda_{0}^{+}=\sum_{i=1}^{\infty}\,
\left(
\begin{array}{c}
2\,i-1\\
i
\end{array}
\right)
(-1)^{i-1}\,\left({B\over
\delta}\right)^{2\,i} ={1\over 2} - {1\over 2 \sqrt{1+\left({\delta\over 2
B}\right)^2}},
\label{lam0p}
\eeq
and
\bea
\Lambda_{1}^{+}&=&\sum_{i=2}^{\infty}\,
\left(
\begin{array}{c}
2\,i-2\\
i
\end{array}
\right)
(-1)^{i-2}\,\left({B\over
\delta}\right)^{2\,i-1},\nnu\\
 &=&
{4 \left({B\over\delta}\right)^3\over  \sqrt{1+\left({2 B\over
\delta}\right)^2} \left(1+\sqrt{1+\left({2 B\over
\delta}\right)^2}\right)^2}\, ,\nnu\\
\label{lam1p}
\eea
which from Eqs.~(\ref{sol3}) lead to $D_{+}={1\over 2}-{1\over 2}
{1\over \sqrt{1+\left({\delta\over 2 B}\right)^2}}$ and $D_{-}=0$.
A similar treatment can be done for
$\Lambda_{i}^-=S_{i-1}^{-}(i-1,i)$ leading to the recursive
relation
\beq
\Lambda^{-}_{i}={B\over
\delta}\left[a^{-}_{i-1}-\Lambda^{-}_{i-1}\right].
\label{lamm}
\eeq
After a lengthy algebra we finally get the instantonic
contribution
\bea
\sum_{i=1}^{\infty} I(i-1,i)&=&\left({1\over
2}-{1\over 2}\,
{1\over \sqrt{1+\left({2 B\over \delta}\right)^2}}\right)\,e^{-E_0\,T}\nnu\\
&-&
e^{-E_1\,T} +
\left({1\over 2}+{1\over 2}\,
{1\over \sqrt{1+\left({2 B\over \delta}\right)^2}}\right)\,e^{-E_2\,T},
\nnu \\
\label{fin2}
\eea
where $E_0$ and $E_2$ as before, while $E_1={\omega_1\over 2}$.

To sum up we finally get our formulae for the energy spectra  for the triple well,
arranged in ascending
order from the ground state,:
\bea
E_0&=& {(\omega_0 + \omega_1)\over 4}-\sqrt{{\delta^2\over 4} +
B^2},\nnu\\
E_1 &=& {\omega_1 \over 2},\nnu\\
E_2&=& {(\omega_0 + \omega_1)\over 4}+\sqrt{{\delta^2\over 4} +
B^2}, \nnu\\
\label{ener2}
\eea
where $B$ for the potential in \eq{pot} is: ${8\over
\sqrt{3\,\pi}}\,\omega^{3/2}\,e^{-{\omega\over4}}$.

Our formulae, Eqs.~(\ref{ener2}), are completely different from
that of Refs.~\cite{lee,casa}. In the limit of large $\omega$, the
spectra tend to the limit $E_0\rightarrow {\omega\over 2}$,
$E_2\rightarrow E_1=\omega$, \ie a singlet and a degenerate
doublet.

To confirm our formulae we resort to numerical solution of
Schr$\ddot{o}$dinger equation for the potential \eq{pot}. In this letter we
present the results for the energy differences in Table~\ref{tab1}
and for the individual energy eigenvalues in Table~\ref{tab2}. The
details of the numerical method can be found in Ref.~\cite{ourdw}.
Moreover our results are consistent with what one could predict
based on simple quantum mechanical approach to the problem
Ref.~\cite{oursim}.
\begin{table}[hbtp]
\begin{tabular}{lllll}
\hline \hline $\omega$ & $\Delta E_{10}^{\mbox{\it num}}$ &
$\Delta E_{10}^{\mbox{\it ins}}$ & $\Delta E_{21}^{\mbox{\it
num}}$ & $\Delta E_{21}^{\mbox{\it ins}}$ \\
\hline \hline
$30$ & $13.67878984$ & $15.00373814$ &$0.004723029$ & $0.003738143$\\
$50$ & $23.77967090$ & $25.00000047$ &$9.100602755\,\times 10^{-7}$ & $4.715381063\,\times 10^{-7}$ \\
$70$ & $33.81102645$ & $35.00000000$ & $1.018592013\,\times 10^{-10}$ & $4.195754855\,\times 10^{-11}$ \\
$90$ &$43.82678307$ & $45.00000000$ & $8.950392428\,\times 10^{-15}$ & $3.552713679\,\times 10^{-15}$ \\
$110$ &$53.83629393$ & $55.00000000$ & $6.844874590\,\times 10^{-19}$ & $2.135638335\,\times 10^{-19}$\\
\hline \hline
\end{tabular}
\caption{\footnotesize The numerically calculated energy
differences ($\Delta E^{\mbox{\it num}}$) against those ($\Delta
E^{\mbox{\it ins}}$) predicted using instantonic  approach. Notice
that $\Delta E_{ij}=E_i - E_j $. }
\label{tab1}
\end{table}
\begin{table}[hbtp]
\begin{tabular}{cccc}
\hline \hline
& $\omega = 30$ & $\omega = 50$ & $\omega = 70$ \\
\hline \hline
$E_0$& $14.178009913879056439$ & $24.211602974742920912$ &$34.22366585434764411968736$ \\
$E_1$& $27.856799752355288816$ & $47.991273870020627020$ &$68.03469230180201792058325$ \\
$E_2$& $27.861522781409804929$ & $47.991274780080902476$ &$68.03469230190387712184547$ \\
\hline \hline
\end{tabular}
\begin{tabular}{ccc}
& $\omega = 90$ & $\omega = 110$ \\
\hline \hline
$E_0$& $44.229945163238886600375384901$ & $54.233802436877662843887532359$ \\
$E_1$& $88.056728233456712807586843449$ & $108.070096366408926121501688376$ \\
$E_2$& $88.056728233456721757979271787$ & $108.070096366408926122186175835$ \\
\hline \hline
\end{tabular}
\caption{\footnotesize The numerically calculated energy
eigenvalues. Numerical results, for large $\omega$, show singlet
and doublet structures ($E_0\rightarrow {\omega\over 2}$,
$E_2\rightarrow E_1=\omega$).} \label{tab2}
\end{table}

In this letter we have emphasized the power of the application of
the instanton method, when a proper treatment is used, and shown
that the dilute gas approximation is sufficient. We have confirmed
our splitting energy formulae numerically. As a final comment we
expect that for a potential well with non-equivalent vacua, the
multiplicity of the splitting is related to the number of
equivalent vacua connected by the discrete symmetry of the
potential.

\section*{Acknowledgement}
This work was supported by research center at college of science,
King Saud University under project number Phys$/1423/02$. A part
of this work was done within the associate scheme of ICTP.

\end{document}